# Perfect Lattice Actions for the Gross-Neveu Model at large $N$ *

W. Bietenholz[1], E. Focht[2] and U.-J. Wiese[1]

[1] Center for Theoretical Physics,
Laboratory for Nuclear Science, and Department of Physics
Massachusetts Institute of Technology (MIT)
Cambridge, Massachusetts 02139, U.S.A.

[2] Gruppe Theorie der Elementarteilchen
Höchstleistungsrechenzentrum (HLRZ)
52425 Jülich, Germany

MIT Preprint, CTP 2356

September 23, 1994

### Abstract

Fixed point actions for free and interacting staggered lattice fermions are constructed by iterating renormalization group transformations. At large $N$ the fixed point action for the Gross-Neveu model is a perfect action in the sense of Hasenfratz and Niedermayer, i.e. cut-off effects are completely eliminated. In particular, the fermionic 1-particle energy spectrum of the lattice theory is identical with the one of the continuum even for arbitrarily small correlation lengths. The cut-off effects of the chiral condensate are eliminated using a perfect operator.

---

*This work is supported in part by funds provided by the U.S. Department of Energy (D.O.E.) under cooperative research agreement DE-FC02-94ER40818.

# 1  Introduction

Cut-off effects are the main source of systematic errors in numerical simulations of lattice field theories. Artifacts due to a finite lattice spacing $a$ vanish in the continuum limit when the correlation length diverges in lattice units. In bosonic theories the lattice artifacts are usually of $O(a^2)$ and in fermionic theories they are of $O(a)$. Hence they go to zero rather slowly as the continuum limit is approached. In practice it is very difficult to work at large correlation lengths mostly because of critical slowing down. When Wilson introduced lattice field theory his idea was to use the renormalization group to map the critical continuum theory — defined at a fixed point of the renormalization group — to a noncritical theory with small correlation length which could then be solved numerically [1]. This requires to approach the renormalized trajectory of theories connected to the fixed point by an infinite number of exact renormalization group steps. Theories on this trajectory have perfect lattice actions — their spectrum is completely free of cut-off effects. Unfortunately, attempts to locate the renormalized trajectory using real space renormalization group techniques have not been successful. Therefore numerical simulations must approach the continuum limit the hard way, going to larger and larger correlation lengths until scaling is finally observed.

The idea of Symanzik's improvement program [2] is to systematically approach the renormalized trajectory by adding irrelevant operators of a given dimension to the standard lattice action. Their coefficients are fixed by demanding that physical quantities become cut-off independent up to a given order in $a$. As the desired order of improvement is increased, more and more operators of higher and higher dimension must be taken into account. In asymptotically free theories the coefficients of the various operators can be computed in weak coupling perturbation theory. The leading contribution defines classical (tree level) improvement, while the higher orders represent quantum ($n$-loop) improvement. For gauge theories Lüscher and Weisz [3] have introduced the concept of on-shell improvement and they have constructed the tree level improved action for Yang-Mills theory to $O(a^2)$. Sheikholeslami and Wohlert [4] have applied similar ideas to QCD and their $O(a)$ tree level improved (so called clover) action is now often used in numerical simulations. In the Gross-Neveu model Wetzel [5] has implemented Symanzik's improvement program in the large $N$ limit.

Recently, Hasenfratz and Niedermayer [6] have taken a more radical approach to eliminate cut-off effects. They realized that perfect actions can be constructed explicitly for asymptotically free field theories. Their method is nonperturbative and it results in the full fixed point action by working with an, in principle, infinite number of irrelevant operators. In practice one is of course limited to a truncated (but still large) set of operators. The fixed point action is a perfect classical action and it should resemble a tree level Symanzik improved action to all orders in $a$. Hasenfratz and Niedermayer also discuss how to derive perfect quantum actions which are located on the renormalized trajectory (and which should correspond to all-loop improvement to all orders in $a$). However, in practice this seems not to be really necessary, at least in the 2-d $O(3)$ nonlinear $\sigma$-model that they investigated numerically. Using the perfect classical action at correlation lengths as small as $5a$ the cut-off effects were smaller than the statistical errors and were hence practically eliminated. At present one does not really understand why the perfect classical action works so well at the quantum level even at these extremely small correlation lengths.

The method of ref.[6] applies to any asymptotically free theory. Of course, we are most interested in QCD, and it is a great challenge to construct its perfect action. However, this needs some preparation and cannot be done in one step. A natural first step is the construction of a perfect action for pure Yang-Mills theory which is in progress [7]. In the second step one would include the quarks. We think that it may be helpful to gain experience with fixed point actions for fermions also in simpler settings. The simplest asymptotically free theory with fermions is the Gross-Neveu model in two dimensions [8]. In the continuum the model with $N$ flavors is defined by the euclidean action

$$S[\bar{\Psi}, \Psi] = \int d^2x \, \{\sum_{i=1}^{N} \bar{\Psi}^i \gamma_\mu \partial_\mu \Psi^i - \frac{G}{2} (\sum_{i=1}^{N} \bar{\Psi}^i \Psi^i)^2\}. \qquad (1.1)$$

For later convenience we rescale the fields to

$$\bar{\chi}^i = \sqrt{G} \bar{\Psi}^i, \;\; \chi^i = \sqrt{G} \Psi^i, \qquad (1.2)$$

and we introduce a real valued auxiliary field $\Phi$ to linearize the 4-Fermi interaction by a Yukawa coupling. Then the action turns into

$$\frac{1}{G} S[\bar{\chi}, \chi, \Phi] = \frac{1}{G} \int d^2x \, \{\sum_{i=1}^{N} \bar{\chi}^i \gamma_\mu \partial_\mu \chi^i + \frac{1}{2} \Phi^2 + \sum_{i=1}^{N} \bar{\chi}^i \chi^i \Phi\}, \qquad (1.3)$$

and the 4-Fermi coupling constant appears as a global prefactor. The model has an $O(2N) \otimes \mathbf{Z}(2)$ chiral symmetry that gets spontaneously broken to $O(2N)$ resulting in a dynamically generated fermion mass $m_f$ [8]. The model is asymptotically free, i.e. in a lattice formulation the continuum limit corresponds to $G \to 0$, and for small values of $G$ one expects

$$am_f(G) \sim \exp(-\pi/G(N-1)). \qquad (1.4)$$

Note that the model with a single flavor ($N = 1$) is identical with the Thirring model which has a vanishing $\beta$-function. Also the $N = 2$ model is special. It has been conjectured that it is equivalent to two decoupled sine-Gordon models. The large $N$ limit is taken such that $g = GN$ is kept fixed. Then asymptotic scaling corresponds to

$$am_f(g) \sim \exp(-\pi/g). \qquad (1.5)$$

The lattice Gross-Neveu model has been studied e.g. in refs.[9, 10, 11] using staggered fermions. Then the number of flavors is even. On the lattice the $O(2N) \otimes \mathbb{Z}(2)$ symmetry of the continuum theory is explicitly broken to $U(N/2)_{e=o} \otimes \mathbb{Z}(2)$. It is essential that the $\mathbb{Z}(2)$ symmetry remains intact, because then the dynamics of its spontaneous breakdown is not obscured by lattice effects.

Here we investigate fixed point actions for the Gross-Neveu model in the large $N$ limit, where the whole calculation can be done analytically. It turns out that at $N = \infty$ the fixed point action — which is the perfect classical action — is in fact also a perfect quantum action. In particular, the energy spectrum of the lattice theory is identical with the one of the continuum even for arbitrarily small correlation lengths. This shows that the success of the method of Hasenfratz and Niedermayer is not limited to the $O(3)$-model, and it hopefully indicates that it will also work for QCD. We also investigate a small field approximation to the fixed point action which turns out not to be a perfect action. This underlines that the nonperturbative approach of ref.[6] is really necessary, and that one should not seek short cuts using small field approximations, although this is usually much simpler than determining the full fixed point action. Moreover, we construct a perfect operator for the chiral condensate to demonstrate that its cut-off effects can also be eliminated. However, this does not happen automatically just by using a perfect action. Then we demonstrate that with the perfect action *asymptotic* scaling sets in earlier than with the standard action. Still, for the perfect action asymptotic scaling is not perfect. Asymptotic scaling is not really what one should ask for. It is an unphysical issue because it involves the bare coupling defined at the cut-off scale. The real issue is scaling of ratios of physical quantities — and scaling is perfect for a perfect action. Finally, we investigate the renormalized trajectory in order to understand why the perfect classical (fixed point) action is even a perfect quantum action. It turns out that this is the case only at $N = \infty$. For finite $N$ one expects cut-off effects at finite correlation lengths even with the perfect classical action. Still, the fact that it becomes perfect at large $N$ may explain why the fixed point action was practically perfect in the $O(3)$ model study of Hasenfratz and Niedermayer. Of course, this argument assumes that the model at $N = 3$ does not deviate significantly from the large $N$ limit.

In section 2 we investigate the fixed point actions for free staggered fermions using a renormalization group transformation suggested by Kalkreuter, Mack and Speh [12] that is consistent with the staggered symmetries. We generalize their renormalization group transformation in a way that allows us to optimize the locality of the fixed point action. In section 3 we turn to the 2-d Gross-Neveu model by switching on a 4-Fermi coupling which is linearized by a Yukawa coupling to an auxiliary scalar field. Blocking both the fermion and the auxiliary scalar field we determine the fixed point action of the coupled system in a small field approximation. Section 4 deals with the large $N$ limit. There we can go beyond the small field approximation for the zero mode of the auxiliary scalar field, which allows us to determine the perfect classical action analytically. We also construct a perfect operator for the chiral condensate. In section 5 the Gross-Neveu model is solved in the large $N$ limit and it is shown that the perfect classical action is in fact also a perfect quantum action. In particular, the cut-off effects of the energy spectrum and of the chiral condensate are completely eliminated. We also investigate the renormalized trajectory to answer the question why the fixed point action is even a perfect quantum action. Finally, we draw some conclusions in section 6.

## 2 Fixed point actions for free staggered fermions

The results of this section have been presented before by two of the authors [13]. A similar study for Wilson fermions was performed in ref.[14]. In the Gross-Neveu model chiral symmetry is spontaneously broken. Since Wilson fermions break chiral symmetry explicitly it is easier to investigate questions of spontaneous chiral symmetry breaking using staggered fermions. They are protected against perturbative radiative mass corrections by a remnant of chiral symmetry. Staggered fermions are also easier to use in numerical simulations because the location of their critical surface is determined by chiral symmetry.

In two dimensions staggered fermions represent two flavors of Dirac fermions in the continuum limit. Staggered fermions can be described by one-component Grassmann variables living on a quadratic lattice with spacing $a/2$. Here we label the staggered fermion variables $\bar\chi_x^i, \chi_x^i$ by two indices $x$ and $i$, where $x$ are the centers of $2 \times 2$ disjoint blocks and $i \in \{1, 2, 3, 4\}$ is a pseudoflavor index that determines the position within a block. The pseudoflavors 1 and 2 correspond to the lower left and right corners while the pseudoflavors 3 and 4 represent the upper left and right corners of the block. The block centers $x$ form a quadratic lattice with spacing $a$. For technical reasons we locate the fermions at the $2 \times 2$ block centers $x$ and not at the sites of the lattice with spacing $a/2$. From now on we set $a = 1$. In general the action of the free theory may be parametrized as

$$S[\bar\chi, \chi] = \sum_{x,y} \sum_{i,j} \bar\chi_x^i \rho_{ij}(x-y) \chi_y^j = \sum_{x,y} \bar\chi_x \rho(x-y) \chi_y. \tag{2.1}$$

Here $\rho(z)$ is a $4 \times 4$ matrix in pseudoflavor space that depends on the difference vector $z = x - y$ between two block centers. Staggered fermions have various symmetries, among them a $U(1)_e \otimes U(1)_o$ remnant of chiral invariance, an analog of charge conjugation, and pseudoflavor transformations that correspond to shift symmetries on the lattice with spacing $1/2$. The symmetries impose constraints on the matrix $\rho(z)$. The $U(1)_e \otimes U(1)_o$ symmetry implies $\rho_{ii}(z) = 0$, $\rho_{14}(z) = \rho_{41}(z) = 0$, $\rho_{23}(z) = \rho_{32}(z) = 0$, charge conjugation leads to $\rho_{ij}(-z) = -\rho_{ji}(z)$, and the shift symmetries give in addition $\rho_{24}(z) = -\rho_{13}(z)$, $\rho_{34}(z) = \rho_{12}(z)$. This leads to the

structure

$$\rho(z) = \begin{pmatrix} 0 & \rho_1(z) & \rho_2(z) & 0 \\ -\rho_1(-z) & 0 & 0 & -\rho_2(z) \\ -\rho_2(-z) & 0 & 0 & \rho_1(z) \\ 0 & \rho_2(-z) & -\rho_1(-z) & 0 \end{pmatrix}. \qquad (2.2)$$

The shift symmetries restrict the matrix elements even further by

$$\rho_\mu(-z) = -\rho_\mu(z + \hat{\mu}), \qquad (2.3)$$

where $\hat{\mu}$ is the unit vector in $\mu$-direction. The standard action is characterized by $\rho(z) = \rho^s(z)$ with

$$\rho^s_\mu(0) = -1, \quad \rho^s_\mu(\hat{\mu}) = 1. \qquad (2.4)$$

All the other values are zero. The actions that we will generate by renormalization group transformations are more general but they are consistent with the symmetry requirements summarized in eqs.(2.2,2.3). Note that we have not introduced a chiral symmetry breaking fermion mass term. The model represents massless free fermions with infinite correlation length and hence it corresponds to a critical lattice field theory.

The standard action represents a particularly simple point on the critical surface but, except for this, not a very interesting one. Here we are interested in fixed points of the renormalization group because this is were the continuum limit of a lattice field theory is defined. To locate the fixed points we start at any point on the critical surface — in this case at the point defined by the standard action — and we iterate renormalization group transformations. The location of a fixed point will in general depend on the renormalization group transformation that we choose, and we will use this freedom to optimize the fixed point action's locality. To be consistent with the symmetries of the problem it is important that the renormalization group transformation respects the pseudoflavor structure of staggered fermions. Kalkreuter, Mack and Speh have proposed a suitable blocking scheme with blocking factor 3 [12]. To respect the staggered fermion symmetries the blocking factor must be odd. The blocked coarse lattice has spacing 3. Hence nine $2 \times 2$ block centers $x$ form a block that is associated with a new $2 \times 2$ block center $x'$ on the coarse lattice. We denote this by $x \in x'$. Each pseudoflavor is blocked individually such that each $\chi^i_x$ on the original fine lattice contributes to exactly one $\chi'^i_{x'}$ on the blocked coarse lattice. The block transformation is illustrated in fig.1. First we apply a $\delta$-function renormalization group transformation such that after one blocking step the effective action $S'[\bar{\chi}', \chi']$ is given by

$$\begin{aligned}\exp(-\frac{1}{G}S'[\bar{\chi}', \chi']) &= \int \mathcal{D}\bar{\chi}\mathcal{D}\chi \exp(-\frac{1}{G}S[\bar{\chi}, \chi]) \\ &\times \prod_{x',i} \delta(\bar{\chi}'^i_{x'} - \frac{b}{3^2}\sum_{x \in x'} \bar{\chi}^i_x)\delta(\chi'^i_{x'} - \frac{b}{3^2}\sum_{x \in x'} \chi^i_x)\end{aligned}$$

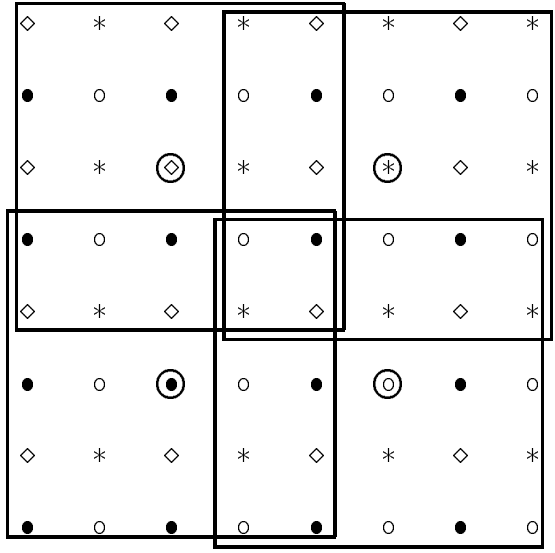

Figure 1: *Blocking of a 2-d lattice consistent with the staggered symmetries. The symbols •, ○, ◇, ∗ represent the pseudoflavors 1, 2, 3, 4 on the lattice with spacing 1/2. Four 3 × 3 blocks of the various pseudoflavors are also shown. Their centers forming the blocked lattice are marked with a large circle.*

$$
\begin{aligned}
&= \int \mathcal{D}\bar{\chi}\mathcal{D}\chi\mathcal{D}\bar{\eta}\mathcal{D}\eta \, \exp(-\frac{1}{G}S[\bar{\chi},\chi]) \\
&\times \exp\{\sum_{x'}\sum_{i}[(\bar{\chi}'^i_{x'} - \frac{b}{3^2}\sum_{x\in x'}\bar{\chi}^i_x)\eta^i_{x'} + \bar{\eta}^i_{x'}(\chi'^i_{x'} - \frac{b}{3^2}\sum_{x\in x'}\chi^i_x)]\}.
\end{aligned}
\tag{2.5}
$$

We have already introduced the 4-Fermi coupling $G$ of the Gross-Neveu model. There due to asymptotic freedom the continuum limit (and hence the critical surface) is at $G \to 0$. Then, as Hasenfratz and Niedermayer pointed out, finding the fixed point action is a classical saddle point problem. For each pseudoflavor we have introduced a coarse lattice auxiliary Grassmann field $\bar{\eta}^i_{x'}$, $\eta^i_{x'}$. The parameter $b$ renormalizes the fermion field. It must be fixed appropriately in order to reach a fixed point of the renormalization group. It is important to note that the renormalization group transformation leaves the partition function — and hence the physics — invariant. This follows immediately when one integrates out the blocked variables

$$
Z' = \int \mathcal{D}\bar{\chi}'\mathcal{D}\chi' \exp(-\frac{1}{G}S'[\bar{\chi}',\chi']) = \int \mathcal{D}\bar{\chi}\mathcal{D}\chi \exp(-\frac{1}{G}S[\bar{\chi},\chi]) = Z. \tag{2.6}
$$

Before we proceed with the blocking step we generalize the renormalization group transformation by introducing a kinetic term for the auxiliary Grassmann field

$$
\begin{aligned}
\exp(-\frac{1}{G}S'[\bar{\chi}',\chi']) &= \int \mathcal{D}\bar{\chi}\mathcal{D}\chi\mathcal{D}\bar{\eta}\mathcal{D}\eta \, \exp(-\frac{1}{G}S[\bar{\chi},\chi]) \\
&\times \exp\{\sum_{x'}\sum_{i}[(\bar{\chi}'^i_{x'} - \frac{b}{3^2}\sum_{x\in x'}\bar{\chi}^i_x)\eta^i_{x'} + \bar{\eta}^i_{x'}(\chi'^i_{x'} - \frac{b}{3^2}\sum_{x\in x'}\chi^i_x)]\}
\end{aligned}
$$

$$\times \quad \exp(-\frac{G}{c}\sum_{x',y'} \bar{\eta}_{x'}\rho^s(\frac{x'-y'}{3})\eta_{y'}). \tag{2.7}$$

Here $\rho^s(z)$ is the matrix defined in eq.(2.4) that characterizes the standard action. The parameter $c$ will be used later to optimize the locality of the fixed point action. It is easy to show that with the extra kinetic term for the fermionic auxiliary field the renormalization group transformation still leaves the partition function invariant. It should be noted that this is not the case for gauge theories. Hence, there one is limited to $c = \infty$. Of course, one could also introduce a mass term for the auxiliary fermionic field. This would, however, break the remnant chiral symmetry and would therefore contradict our strategy. This is in contrast to Wilson fermions where chiral symmetry is completely broken already by the action. In that case it is natural to introduce additional chiral breaking via the renormalization group transformation [14].

To perform the renormalization group step it is useful to go to momentum space. Then all variables are replaced by their Fourier transforms, e.g.

$$\rho(z) = \frac{1}{(2\pi)^2}\int_B d^2k\, \rho(k)\exp(ikz). \tag{2.8}$$

Here we integrate over the Brillouin zone $B = ]-\pi,\pi]^2$. The quantities defined on the blocked lattice are integrated over a smaller Brillouin zone $B' = ]-\pi/3,\pi/3]^2$. In momentum space eq.(2.7) takes the form

$$\begin{aligned}\exp(-\frac{1}{G}S'[\bar{\chi}',\chi']) &= \int \mathcal{D}\bar{\chi}\mathcal{D}\chi\mathcal{D}\bar{\eta}\mathcal{D}\eta \exp\{-\frac{1}{G}\frac{1}{(2\pi)^2}\int_B d^2k\, \bar{\chi}(-k)\rho(k)\chi(k) \\ &+ \left(\frac{3}{2\pi}\right)^2 \int_{B'} d^2k\, [\bar{\psi}'(-k)\eta(k) + \bar{\eta}(-k)\psi'(k)] \\ &- \frac{G}{c}\left(\frac{3}{2\pi}\right)^2 \int_{B'} d^2k\, \bar{\eta}(-k)\rho^s(3k)\eta(k)\}. \end{aligned} \tag{2.9}$$

We have defined

$$\begin{aligned}\psi'(k) &= \chi'(k) - \frac{b}{3^2}\sum_l \Pi(k+\frac{2\pi l}{3})D(k+\frac{2\pi l}{3})\chi(k+\frac{2\pi l}{3}), \\ \bar{\psi}'(-k) &= \bar{\chi}'(-k) - \frac{b}{3^2}\sum_l \bar{\chi}(-k-\frac{2\pi l}{3})D(-k-\frac{2\pi l}{3})\Pi(k+\frac{2\pi l}{3}). \end{aligned} \tag{2.10}$$

The sums extend over integer vectors with components $l_\mu \in \{1,2,3\}$. We have also introduced a matrix $D(k)$ diagonal in pseudoflavor space that is given by

$$\begin{aligned}D_{11}(k) &= \exp\frac{i}{2}(-k_1-k_2), \quad D_{22}(k) = \exp\frac{i}{2}(k_1-k_2), \\ D_{33}(k) &= \exp\frac{i}{2}(-k_1+k_2), \quad D_{44}(k) = \exp\frac{i}{2}(k_1+k_2).\end{aligned} \tag{2.11}$$

It arises because in our notation the fermionic variables are located at the $2 \times 2$ block centers $x$ forming the lattice with spacing 1, and not at the sites of the lattice with spacing $1/2$. Finally, the $3 \times 3$ block average gives rise to the function $\Pi(k) = \prod_\nu \frac{1}{3}(1 + 2\cos k_\nu)$. To obtain the effective action we first integrate out the fermionic auxiliary field

$$\exp(-\frac{1}{G}S'[\bar\chi', \chi']) = \int \mathcal{D}\bar\chi \mathcal{D}\chi \exp\{-\frac{1}{G}\frac{1}{(2\pi)^2}\int_B d^2k\, \bar\chi(-k)\rho(k)\chi(k)$$
$$+ \frac{c}{G}\left(\frac{3}{2\pi}\right)^2 \int_{B'} d^2k\, \bar\psi'(-k)\alpha^s(3k)\psi'(k)\}, \qquad (2.12)$$

where $\alpha^s(k) = \rho^s(k)^{-1}$. Next we solve the saddle point problem, i.e. we derive classical equations of motion for the fermionic variables by varying the exponent in eq.(2.12) with respect to $\bar\chi(-k)$. Keeping in mind that $\bar\psi'(-k)$ also depends on $\bar\chi(-k)$ one finds the classical solution

$$\chi_c(k) = -bc\,\alpha(k)D(-k)\Pi(k)\alpha^s(3k)\psi'(k), \qquad (2.13)$$

where $\alpha(k) = \rho(k)^{-1}$. Multiplying eq.(2.13) by $D(k)\Pi(k)$ and summing it over the momenta $k + 2\pi l/3$ with $l_\mu \in \{1, 2, 3\}$ one obtains

$$\psi'(k) = \chi'(k) + c\,\omega(k)\alpha^s(3k)\psi'(k), \qquad (2.14)$$

with the matrix

$$\omega(k) = \frac{b^2}{3^2}\sum_l D(k + \frac{2\pi l}{3})\alpha(k + \frac{2\pi l}{3})D(-k - \frac{2\pi l}{3})\Pi(k + \frac{2\pi l}{3})^2. \qquad (2.15)$$

¿From eq.(2.14) we infer

$$\psi'(k) = (\mathbb{1} - c\,\omega(k)\alpha^s(3k))^{-1}\chi'(k). \qquad (2.16)$$

Similar relations hold for $\bar\chi_c(-k)$ and $\bar\psi'(-k)$. Inserting these as well as eq.(2.13) and eq.(2.16) back into the exponent of eq.(2.12) one obtains after some manipulations

$$S'[\bar\chi', \chi'] = -c\left(\frac{3}{2\pi}\right)^2 \int_{B'} d^2k\, \bar\chi'(-k)\alpha^s(3k)(\mathbb{1} - c\,\omega(k)\alpha^s(3k))^{-1}\chi'(k). \qquad (2.17)$$

Hence one identifies

$$\alpha'(3k)^{-1} = \rho'(3k) = -c\,\alpha^s(3k)(\mathbb{1} - c\,\omega(k)\alpha^s(3k))^{-1}. \qquad (2.18)$$

Using eqs.(2.16,2.18) we rewrite eq.(2.13) as

$$\chi_c(k) = b\,\alpha(k)D(-k)\Pi(k)\alpha'(3k)^{-1}\chi'(k). \qquad (2.19)$$

Similarly, one derives

$$\bar\chi_c(-k) = b\,\bar\chi'(-k)\alpha'(3k)^{-1}\Pi(k)D(k)\alpha(k). \qquad (2.20)$$

These equations will be useful later because they allow to determine the fixed point action for the interacting theory in a simple manner. Now we return to eq.(2.18) which we rewrite as $\alpha'(k) = \omega(\frac{k}{3}) - \frac{1}{c}\rho^s(k)$ such that

$$\alpha'(k) = \frac{b^2}{3^2} \sum_l D(\frac{k+2\pi l}{3})\alpha(\frac{k+2\pi l}{3})D(-\frac{k+2\pi l}{3})\Pi(\frac{k+2\pi l}{3})^2 - \frac{1}{c}\rho^s(k). \quad (2.21)$$

To eliminate the $D$-factors we introduce the matrix $\tilde{\alpha}(k) = D(-k)^{1/2}\alpha(k)D(k)^{1/2}$ which has the structure [1]

$$\tilde{\alpha}(k) = \begin{pmatrix} 0 & \tilde{\alpha}_1(k) & \tilde{\alpha}_2(k) & 0 \\ \tilde{\alpha}_1(k) & 0 & 0 & -\tilde{\alpha}_2(k) \\ \tilde{\alpha}_2(k) & 0 & 0 & \tilde{\alpha}_1(k) \\ 0 & -\tilde{\alpha}_2(k) & \tilde{\alpha}_1(k) & 0 \end{pmatrix}. \quad (2.22)$$

In coordinate space the multiplication with $D(k)^{1/2}$ corresponds to a pseudoflavor dependent shift of the fermionic variables by $\pm 1/4$ of the lattice spacing in each direction. This shifts the fermions away from the centers $x$ of the $2 \times 2$ blocks (that form the lattice with spacing 1) to the sites of the lattice with spacing $1/2$ that is illustrated in fig.1. Generally, all quantities with tildes (there will be more defined later) refer to these shifted fermions. Eq.(2.21) turns into the recursion relation

$$\tilde{\alpha}'_\mu(k) = \frac{b^2}{3^2} \sum_l \tilde{\alpha}_\mu(\frac{k+2\pi l}{3})(-1)^{l_\mu} \prod_\nu \left(\frac{2\sin(k_\nu/2)}{6\sin((k_\nu+2\pi l_\nu)/6)}\right)^2 + \frac{1}{c}i\hat{k}_\mu. \quad (2.23)$$

We define $\tilde{\rho}^s(k) = D(-k)^{1/2}\rho^s(k)D(k)^{1/2}$ and we have used $\tilde{\rho}^s_\mu(k) = -i\hat{k}_\mu$ with $\hat{k}_\mu = 2\sin(k_\mu/2)$ as well as $1 + 2\cos(k_\nu) = \sin(3k_\nu/2)/\sin(k_\nu/2)$. The sign factor $(-1)^{l_\mu}$ in eq.(2.23) arises after the $D$-factors have canceled. Now we iterate eq.(2.23) and after $n$ renormalization group steps we find

$$\tilde{\alpha}^{(n)}_\mu(k) = \left(\frac{b^2}{3^2}\right)^n \sum_l \tilde{\alpha}_\mu(\frac{k+2\pi l}{3^n})(-1)^{l_\mu} \prod_\nu \left(\frac{2\sin(k_\nu/2)}{3^n 2\sin((k_\nu+2\pi l_\nu)/3^n 2)}\right)^2$$
$$+ \frac{1-(b^2/3^3)^n}{c(1-b^2/3^3)} i\hat{k}_\mu, \quad (2.24)$$

with $l_\mu \in \{1, 2, ..., 3^n\}$. In the limit $n \to \infty$ only the small $k$ behavior of $\tilde{\alpha}_\mu(k)$ close to its pole in $B$ is important. Here we have $\tilde{\alpha}_\mu(k) = i\hat{k}_\mu/\hat{k}^2 \sim ik_\mu/k^2$ such that after an infinite number of renormalization group steps

$$\tilde{\alpha}^*_\mu(k) = \lim_{n \to \infty} \left(\frac{b^2}{3^2}\right)^n i \sum_{l \in \mathbb{Z}^2} \frac{3^n(k_\mu+2\pi l_\mu)}{(k+2\pi l)^2}(-1)^{l_\mu} \prod_\nu \left(\frac{2\sin(k_\nu/2)}{k_\nu+2\pi l_\nu}\right)^2$$
$$+ \frac{1-(b^2/3^3)^n}{c(1-b^2/3^3)} i\hat{k}_\mu. \quad (2.25)$$

---

[1]In ref.[13] we forgot to put the tilde on $\alpha(k)$.

The limit is finite — i.e. a nontrivial fixed point is reached — only if $(b^2/3^2)^n 3^n = 1$ which implies $b = \sqrt{3}$. This is what one would expect on dimensional grounds. The canonical dimension of a fermion field in two dimensions is $d_\chi = 1/2$. In order to renormalize the fermion field by the appropriate amount, $b$ should be given by $3^{d_\chi}$. At the fixed point one obtains

$$\tilde{\alpha}_\mu^*(k) = i \sum_{l \in \mathbb{Z}^2} \frac{k_\mu + 2\pi l_\mu}{(k+2\pi l)^2} (-1)^{l_\mu} \prod_\nu \left( \frac{2\sin(k_\nu/2)}{k_\nu + 2\pi l_\nu} \right)^2 + \frac{9}{8c} i\hat{k}_\mu. \qquad (2.26)$$

Now the parameter $c$ is used to optimize the locality of the fixed point action. For this purpose we consider field configurations that are constant in the 2-direction, i.e. in momentum space they have $k_2 = 0$. For these effectively 1-dimensional configurations it is sufficient to consider $\tilde{\alpha}_\mu^*(k_1, 0)$. One finds $\tilde{\alpha}_2^*(k_1, 0) = 0$ and

$$\tilde{\alpha}_1^*(k_1, 0) = i \sum_{l_1 \in \mathbb{Z}} \frac{1}{k_1 + 2\pi l_1} (-1)^{l_1} \left( \frac{2\sin(k_1/2)}{k_1 + 2\pi l_1} \right)^2 + \frac{9}{8c} i\hat{k}_1 = i\left(\frac{1}{\hat{k}_1} - \frac{1}{8}\hat{k}_1 + \frac{9}{8c}\hat{k}_1\right). \qquad (2.27)$$

For $c = 9$ we have $\tilde{\rho}_1^*(k_1, 0) = \tilde{\alpha}_1^*(k_1, 0)^{-1} = -i\hat{k}_1$ which corresponds exactly to the standard action with nearest neighbor couplings. This is maximally local. For general 2-dimensional configurations the fixed point action with $c = 9$ not only contains nearest neighbor interactions. However, as shown in fig.2 the corresponding function $\tilde{\rho}_1^*(z)$ in coordinate space is still very local. Some values of $\tilde{\rho}_1^*(z)$ are given in table 1. Being a quantity with a tilde $\tilde{\rho}_1^*(z_1, z_2)$ refers to the shifted fermions located at the sites of the lattice with spacing $1/2$. It describes the coupling between the pseudoflavors 1 and 2, which are separated by a half-odd integer multiple of the lattice spacing in the 1-direction and by an integer multiple of the lattice spacing in the 2-direction. Hence $z_1$ is a half-odd integer while $z_2$ is an integer. This also explains the boundary conditions of the corresponding quantity $\tilde{\rho}_1^*(k)$ (and similarly $\tilde{\alpha}_1^*(k)$) in momentum space. In the 1-direction both are antiperiodic over the Brillouin zone due to the corresponding half-odd integer differences in coordinate space, and they are periodic in the 2-direction because there the differences are integers.

# 3   Fixed point actions for the Gross-Neveu model in the small field approximation

In this section we turn to the Gross-Neveu model [8], i.e. we switch on a 4-Fermi interaction which is linearized by a Yukawa coupling to a real valued auxiliary scalar field. The interaction breaks the $U(1)_e \otimes U(1)_o$ chiral symmetry of the free theory explicitly down to $U(1)_{e=o} \otimes \mathbb{Z}(2)$. Still, the $\mathbb{Z}(2)$ symmetry is sufficient to protect the fermions against perturbative radiative mass corrections. However, the interaction is strong enough to break the **Z**(2) symmetry spontaneously which results in a nonperturbatively generated fermion mass. The corresponding order parameter is the chiral condensate.

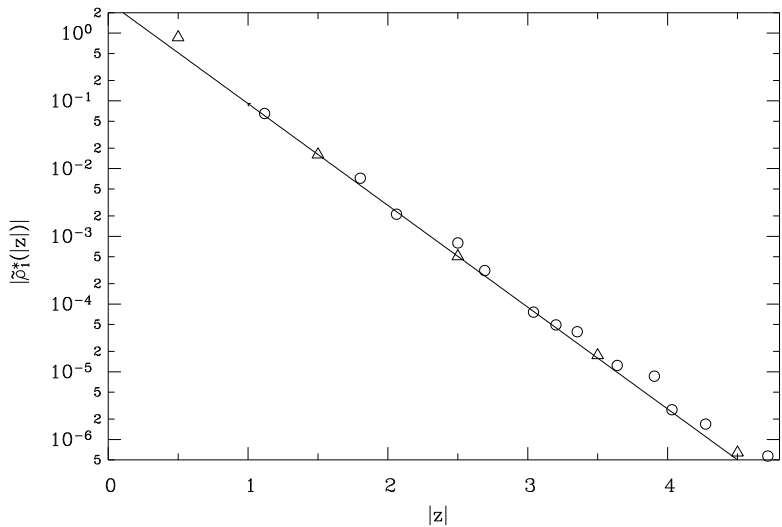

Figure 2: *Exponential decay of the coupling $|\tilde{\rho}_1^*(|z|)|$ as a function of $|z|$ for the optimally local fixed point action with $c = 9$. The triangles correspond to difference vectors parallel to the lattice axes. The straight line is a fit to the triangles resulting in $|\tilde{\rho}_1^*(z_1, 0)| \propto \exp(-3.46|z_1|)$.*

As it was shown in ref.[10] the auxiliary scalar field $\Phi$ should be placed at the centers of the plaquettes of the lattice with spacing $1/2$. This geometry — consistent with the so-called hypercubic Yukawa coupling — is illustrated in fig.3. We now parametrize the action as

$$\frac{1}{G}S[\bar{\chi}, \chi, \Phi] = \frac{1}{G}\{\sum_{x,y} \bar{\chi}_x \rho(x-y)\chi_y + \sum_z \frac{1}{2}\Phi_z^2 + \sum_{x,y,z} \bar{\chi}_x \sigma(x-z, y-z)\chi_y \Phi_z\}. \quad (3.1)$$

Here $\sigma(v, w)$ is a $4 \times 4$ matrix in pseudoflavor space that depends on the difference vectors $v = x - z$ and $w = y - z$ between the fermionic and auxiliary scalar variables. Again the symmetries of the problem impose constraints. The remnant chiral symmetry implies $\sigma_{12}(v,w) = \sigma_{21}(v,w) = 0$, $\sigma_{13}(v,w) = \sigma_{31}(v,w) = 0$, $\sigma_{24}(v,w) = \sigma_{42}(v,w) = 0$, $\sigma_{34}(v,w) = \sigma_{43}(v,w) = 0$, charge conjugation leads to $\sigma_{ij}(w,v) = \sigma_{ji}(v,w)$, and the shift symmetries impose further constraints that we don't write down explicitly.

To determine the fixed point action for the interacting theory one must also define a renormalization group transformation for the auxiliary scalar field. Since the scalar field does not carry a pseudoflavor index one could form ordinary $3 \times 3$ blocks of neighboring variables. Here we prefer to work with scalar field blocks that are of the same kind as the fermionic ones. This is illustrated in fig.3. One may even argue that a scalar pseudoflavor is defined by the surrounding fermionic variables on the corresponding plaquette. Blocking schemes that respect this structure are limited to blocking factors 5, 9, 13, etc. We have checked explicitly that the resulting fixed point actions are identical with the ones we derive here.

| $z_1$ | $z_2$ | $\tilde{\rho}_1^*(z_1, z_2)$ |
|---|---|---|
| 0.5 | 0.0 | 0.8656403 |
| 1.5 | 0.0 | 0.0161216 |
| 2.5 | 0.0 | 0.0005065 |
| 3.5 | 0.0 | 0.0000174 |
| 0.5 | 1.0 | 0.0649779 |
| 1.5 | 1.0 | -0.0072205 |
| 2.5 | 1.0 | -0.0003117 |
| 0.5 | 2.0 | 0.0021230 |
| 1.5 | 2.0 | -0.0007996 |
| 0.5 | 3.0 | 0.0000788 |

Table 1: *Some values of the function $\tilde{\rho}_1^*(z)$ for the optimally local fixed point action with $c = 9$.*

First, we determine the fixed point action for the free auxiliary field. The renormalization group step is then defined as

$$\exp(-\frac{1}{G}S'[\Phi']) = \int \mathcal{D}\Phi \exp(-\frac{1}{G}\sum_z \frac{1}{2}\Phi_z^2) \exp(-\frac{\alpha}{2G}\sum_{z'}(\Phi_{z'} - \frac{\beta}{3^2}\sum_{z \in z'}\Phi_z)^2). \quad (3.2)$$

Finding the fixed point is much simpler here than it was in the fermionic case. In fact, one need not even go to momentum space. We simply quote the result

$$S'[\Phi'] = \sum_{z'} \frac{1}{2}(\alpha - \frac{\alpha^2 \beta^2}{3^2 + \alpha\beta^2})\Phi_{z'}^{'2}. \quad (3.3)$$

Thus the fixed point condition reads

$$\alpha - \frac{\alpha^2 \beta^2}{3^2 + \alpha\beta^2} = 1. \quad (3.4)$$

The auxiliary field remains auxiliary after the renormalization group step, in particular, no kinetic term is generated. Thus, the dimension of $\Phi$ is $d_\Phi = 1$ and not zero as for a genuine dynamical scalar field in two dimensions. Hence, we expect $\beta = 3^{d_\Phi} = 3$. Then eq.(3.4) implies $\alpha = \infty$, i.e. we are restricted to a $\delta$-function renormalization group transformation for the auxiliary field. We will see later that this is indeed the case. For the moment we leave $\alpha$ and $\beta$ undetermined. When we come to the interacting theory we need the relation between the classical solution $\Phi_c$ and the blocked field $\Phi'$ in momentum space which is given by

$$\Phi_c(k) = \frac{3^2 \alpha\beta}{3^2 + \alpha\beta^2}\Pi(k)\Phi'(k). \tag{3.5}$$

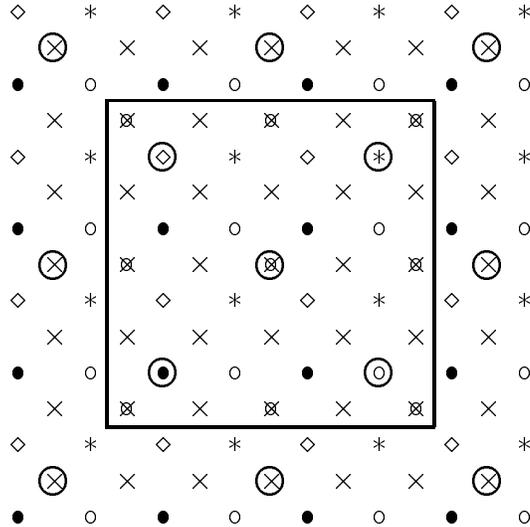

Figure 3: *Geometry of the fermionic and auxiliary scalar variables in the Gross-Neveu model with hypercubic Yukawa coupling. The crosses represent the auxiliary scalar field. A $3\times 3$ block of auxiliary field variables is also shown. The nine crosses that contribute to the block variable are marked with a small circle.*

It should be noted that the fixed point action for the free auxiliary field does not correspond to a critical theory. In fact, the correlation length of the auxiliary field vanishes instead of being infinite.

In the next step we determine the fixed point action for the interacting theory in the small field approximation, i.e. we assume that $\Phi$ is small and we restrict ourselves to $O(\Phi)$ respectively $O(\Phi')$. Up to now we have determined the blocked action as

$$S'[\bar{\chi}', \chi', \Phi'] = S'[\bar{\chi}', \chi'] + S'[\Phi'] + O(\Phi'). \tag{3.6}$$

Since we are at $G \to 0$ the value of $S'[\bar{\chi}', \chi']$ was determined by inserting the solutions $\bar{\chi}_c, \chi_c$ of classical equations of motion (2.19, 2.20) into the exponent of eq.(2.12). We write symbolically

$$S'[\bar{\chi}', \chi'] = E[\bar{\chi}_c, \chi_c]. \tag{3.7}$$

Similarly $S'[\Phi'] = E[\Phi_c]$. The exponent we are dealing with now contains the Yukawa interaction as an additional term of $O(\Phi)$

$$E[\bar{\chi}, \chi, \Phi] = E[\bar{\chi}, \chi] + E[\Phi] + Y[\bar{\chi}, \chi, \Phi]. \tag{3.8}$$

This term changes the classical equations of motion and hence their solutions at $O(\Phi')$ to $\bar{\chi}_c + \delta\bar{\chi}_c + O(\Phi'^2)$, $\chi_c + \delta\chi_c + O(\Phi'^2)$, and $\Phi_c + \delta\Phi_c + O(\Phi'^2)$, where $\delta\bar{\chi}_c$, $\delta \chi_c$ and $\delta \Phi_c$ are of $O(\Phi')$. To determine the blocked action to $O(\Phi')$ one must insert the new solutions into the new exponent

$$\begin{aligned} S'[\bar{\chi}', \chi', \Phi'] &= E[\bar{\chi}_c + \delta\bar{\chi}_c, \chi_c + \delta\chi_c, \Phi_c + \delta\Phi_c] + O(\Phi'^2) \\ &= E[\bar{\chi}_c + \delta\bar{\chi}_c, \chi_c + \delta\chi_c] + E[\Phi_c + \delta\Phi_c] + Y[\bar{\chi}_c, \chi_c, \Phi_c] + O(\Phi'^2) \\ &= E[\bar{\chi}_c, \chi_c] + E[\Phi_c] + Y[\bar{\chi}_c, \chi_c, \Phi_c] + O(\Phi'^2) \\ &= S'[\bar{\chi}', \chi'] + S'[\Phi'] + Y[\bar{\chi}_c, \chi_c, \Phi_c] + O(\Phi'^2). \end{aligned} \quad (3.9)$$

The essential observation is that $\bar{\chi}_c$, $\chi_c$ and $\Phi_c$ are solutions of classical equations of motion, i.e. extrema of the corresponding exponents. This implies $E[\bar{\chi}_c + \delta\bar{\chi}_c, \chi_c + \delta\chi_c] = E[\bar{\chi}_c, \chi_c] + O(\Phi'^2)$ and $E[\Phi_c + \delta\Phi_c] = E[\Phi_c] + O(\Phi'^2)$. Consequently, to determine the blocked action to $O(\Phi')$ at $G \to 0$ it is sufficient to insert the old solutions of the equations of motion into the Yukawa term. For this purpose we again go to momentum space. Then the Yukawa term takes the form

$$Y[\bar{\chi}, \chi, \Phi] = \frac{1}{(4\pi)^4} \int_{\bar{B}^2} d^2p \, d^2q \, \bar{\chi}(-p) \sigma(p,q) \chi(-q) \Phi(p+q). \quad (3.10)$$

The integration of both momenta $p$ and $q$ extends over a large Brillouin zone $\bar{B} = ]-2\pi, 2\pi]^2$ because the auxiliary scalar field lives on a lattice with spacing $1/2$. To obtain the Yukawa term after one renormalization group step we simply insert the old solutions of the classical equations of motion from eqs.(2.19,2.20,3.5) into this expression and we find

$$\begin{aligned} Y'[\bar{\chi}', \chi', \Phi'] &= Y[\bar{\chi}_c, \chi_c, \Phi_c] \\ &= b^2 \frac{3^2 \alpha\beta}{3^2 + \alpha\beta^2} \frac{1}{(4\pi)^4} \int_{\bar{B}^2} d^2p \, d^2q \, \bar{\chi}'(-p) \alpha'(3p)^{-1} \Pi(p) D(p) \alpha(p) \\ &\quad \times \sigma(p,q) \alpha(-q) D(q) \Pi(q) \alpha'(-3q)^{-1} \chi'(-q) \Pi(p+q) \Phi'(p+q) \\ &= \left(\frac{3}{4\pi}\right)^4 \int_{\bar{B}'^2} d^2p \, d^2q \, \bar{\chi}'(-p) \sigma'(3p, 3q) \chi'(-q) \Phi'(p+q). \end{aligned} \quad (3.11)$$

After the renormalization group step the integration extends over the smaller Brillouin zone $\bar{B}' = ]-2\pi/3, 2\pi/3]^2$. One reads off the recursion relation

$$\begin{aligned} \sigma'(p,q) &= b^2 \frac{3^2 \alpha\beta}{3^2 + \alpha\beta^2} \frac{1}{3^4} \sum_{l,m} \alpha'(p)^{-1} D(\frac{p+4\pi l}{3}) \alpha(\frac{p+4\pi l}{3}) \\ &\quad \times \sigma(\frac{p+4\pi l}{3}, \frac{q+4\pi m}{3}) \alpha(-\frac{q+4\pi m}{3}) D(\frac{q+4\pi m}{3}) \alpha'(-q)^{-1} \\ &\quad \times \Pi(\frac{p+4\pi l}{3}) \Pi(\frac{q+4\pi m}{3}) \Pi(\frac{p+4\pi l + q + 4\pi m}{3}). \end{aligned} \quad (3.12)$$

Introducing $\tilde{\sigma}(p,q) = D(-p)^{1/2} \alpha(p) \sigma(p,q) \alpha(-q) D(-q)^{1/2}$, eq.(3.12) turns into the recursion relation

$$\tilde{\sigma}'(p,q) = b^2 \frac{\alpha\beta}{3^2 + \alpha\beta^2} \frac{1}{3^2} \sum_{l,m} \tilde{\sigma}(\frac{p+4\pi l}{3}, \frac{q+4\pi m}{3})$$

$$\times \prod_\nu (-1)^{l_\nu+m_\nu} \frac{2\sin(p_\nu/2)}{6\sin((p_\nu+4\pi l_\nu)/6)} \frac{2\sin(q_\nu/2)}{6\sin((q_\nu+4\pi m_\nu)/6)}$$

$$\times \frac{2\sin((p_\nu+q_\nu)/2)}{6\sin((p_\nu+4\pi l_\nu+q_\nu+4\pi m_\nu)/6)}. \tag{3.13}$$

Iterating this we obtain after $n$ renormalization group steps

$$\tilde{\sigma}^{(n)}(p,q) = \left(b^2 \frac{\alpha\beta}{3^2+\alpha\beta^2} \frac{1}{3^2}\right)^n \sum_{l,m} \tilde{\sigma}(\frac{p+4\pi l}{3^n}, \frac{q+4\pi m}{3^n})$$

$$\times \prod_\nu (-1)^{l_\nu+m_\nu} \frac{2\sin(p_\nu/2)}{3^n 2\sin((p_\nu+4\pi l_\nu)/3^n 2)} \frac{2\sin(q_\nu/2)}{3^n 2\sin((q_\nu+4\pi m_\nu)/3^n 2)}$$

$$\times \frac{2\sin((p_\nu+q_\nu)/2)}{3^n 2\sin((p_\nu+4\pi l_\nu+q_\nu+4\pi m_\nu)/3^n 2)}. \tag{3.14}$$

The standard action is characterized by

$$\tilde{\sigma}(p,q) = \begin{pmatrix} \tilde{\sigma}_0(p,q) & 0 & 0 & -\tilde{\sigma}_3(p,q) \\ 0 & \tilde{\sigma}_0(p,q) & \tilde{\sigma}_3(p,q) & 0 \\ 0 & -\tilde{\sigma}_3(p,q) & \tilde{\sigma}_0(p,q) & 0 \\ \tilde{\sigma}_3(p,q) & 0 & 0 & \tilde{\sigma}_0(p,q) \end{pmatrix} \tag{3.15}$$

with

$$\tilde{\sigma}_0(p,q) = \frac{\hat{p}_\mu \hat{q}_\mu}{\hat{p}^2 \hat{q}^2} \prod_\nu \cos(\frac{p_\nu+q_\nu}{4}),$$

$$\tilde{\sigma}_3(p,q) = \frac{\epsilon_{\mu\rho} \hat{p}_\mu \hat{q}_\rho}{\hat{p}^2 \hat{q}^2} \prod_\nu \cos(\frac{p_\nu+q_\nu}{4}). \tag{3.16}$$

Here $\epsilon_{\mu\rho}$ is the antisymmetric tensor in two dimensions. In the limit of infinitely many renormalization group steps only the behavior of $\tilde{\sigma}(p,q)$ close to its poles matters. The poles that contribute to the fixed point action are at $p = q = 2\pi i$ with $i_\mu \in \{0,1\}$. After infinitely many steps one obtains

$$\tilde{\sigma}_0^*(p,q) = \lim_{n\to\infty} \left(b^2 \frac{\alpha\beta}{3^2+\alpha\beta^2}\right)^n \sum_{l,m} \sum_i \frac{(p_\mu+4\pi l_\mu+2\pi i_\mu)(q_\mu+4\pi m_\mu+2\pi i_\mu)}{(p+4\pi l+2\pi i)^2 (q+4\pi m+2\pi i)^2}$$

$$\times \prod_\nu (-1)^{l_\nu+m_\nu+i_\nu} \frac{2\sin(p_\nu/2)}{p_\nu+4\pi l_\nu+2\pi i_\nu} \frac{2\sin(q_\nu/2)}{q_\nu+4\pi m_\nu+2\pi i_\nu}$$

$$\times \frac{2\sin((p_\nu+q_\nu)/2)}{p_\nu+4\pi l_\nu+q_\nu+4\pi m_\nu+4\pi i_\nu}. \tag{3.17}$$

A nontrivial fixed point is reached only if

$$b^2 \frac{\alpha\beta}{3^2+\alpha\beta^2} = 1. \tag{3.18}$$

Using $b^2 = 3$ together with the fixed point condition eq.(3.4) for the free auxiliary scalar field this implies $\alpha = \infty$ and $\beta = 3$. Hence, one is limited to a $\delta$-function renormalization group transformation for the auxiliary field. The canonical dimension of the auxiliary field is $d_\Phi = 1$ consistent with $\beta = 3^{d_\Phi} = 3$ as discussed earlier. Only for this value the Yukawa coupling in two dimensions is a marginal operator. In particular, for a genuine dynamical scalar field (which would have $d_\Phi = 0$ in two dimensions) the Yukawa coupling would be irrelevant. In fact, the Yukawa coupling to the auxiliary scalar field is weakly relevant. This effect is, however, invisible in the leading order of the weak field approximation. All this is expected because just in two dimensions the Gross-Neveu model is asymptotically free. At the fixed point we finally obtain

$$\begin{aligned}
\tilde{\sigma}_0^*(p,q) &= \sum_{l,m \in \mathbb{Z}^2} \sum_i \frac{(p_\mu + 4\pi l_\mu + 2\pi i_\mu)(q_\mu + 4\pi m_\mu + 2\pi i_\mu)}{(p + 4\pi l + 2\pi i)^2 (q + 4\pi m + 2\pi i)^2} \\
&\quad \times \prod_\nu (-1)^{l_\nu + m_\nu + i_\nu} \frac{2\sin(p_\nu/2)}{p_\nu + 4\pi l_\nu + 2\pi i_\nu} \frac{2\sin(q_\nu/2)}{q_\nu + 4\pi m_\nu + 2\pi i_\nu} \\
&\quad \times \frac{2\sin((p_\nu + q_\nu)/2)}{p_\nu + 4\pi l_\nu + q_\nu + 4\pi m_\nu + 4\pi i_\nu}, \\
\tilde{\sigma}_3^*(p,q) &= \sum_{l,m \in \mathbb{Z}^2} \sum_i \frac{\epsilon_{\mu\rho}(p_\mu + 4\pi l_\mu + 2\pi i_\mu)(q_\rho + 4\pi m_\rho + 2\pi i_\rho)}{(p + 4\pi l + 2\pi i)^2 (q + 4\pi m + 2\pi i)^2} \\
&\quad \times \prod_\nu (-1)^{l_\nu + m_\nu} \frac{2\sin(p_\nu/2)}{p_\nu + 4\pi l_\nu + 2\pi i_\nu} \frac{2\sin(q_\nu/2)}{q_\nu + 4\pi m_\nu + 2\pi i_\nu} \\
&\quad \times \frac{2\sin((p_\nu + q_\nu)/2)}{p_\nu + 4\pi l_\nu + q_\nu + 4\pi m_\nu + 4\pi i_\nu}.
\end{aligned} \quad (3.19)$$

These functions are $4\pi$-antiperiodic in both momenta over the large Brillouin zone $\bar{B}$. Being quantities with a tilde, they refer to the shifted fermions living on the sites of the lattice with spacing $1/2$. Then the fermionic and auxiliary scalar degrees of freedom are separated by multiples of $1/4$ of the lattice spacing. In momentum space this would imply $\pm i$-periodic boundary conditions over the ordinary Brillouin zone $B$, and it implies antiperiodic boundary conditions over the larger Brillouin zone $\bar{B}$.

# 4  Fixed point actions for the Gross-Neveu model at large $N$

In the $O(3)$ model Hasenfratz and Niedermayer have determined the full (large field) fixed point action using a numerical minimization procedure on a multigrid [6]. Still, in their calculation the small field approximation was essential because it allowed them to optimize the fixed point action's locality analytically. In general it will be impossible to go beyond the small field approximation using analytic methods. The only exceptions we are aware of are free field theories and theories in the large $N$ limit. Here we consider the Gross-Neveu model, but the method should work just as well e.g. for $O(N)$ and $CP(N)$ models. For $N = \infty$ the fixed point of the $O(N)$ model has been studied in ref.[15]. For a large number $N$ of flavors the dynamics of the Gross-Neveu model greatly simplifies, because then the auxiliary scalar field is dominated by a constant mean field $\Phi_0$, whose value is determined by the minimum of its effective potential. In fact, in the large $N$ limit only the constant zero mode can have a large value while the amplitudes of the nonzero modes are suppressed by powers of $1/N$ and are thus small. Hence, the small field approximation is justified for the nonzero modes, and only the zero mode needs special treatment. First we concentrate entirely on the zero mode assuming that the auxiliary scalar field is a constant $\Phi_0$. Later we include the nonzero modes using the small field approximation to leading order. Since a constant field remains constant after $\delta$-function blocking, the renormalization group step for the auxiliary field reduces to

$$\Phi_0' = \beta \Phi_0 = 3\Phi_0. \tag{4.1}$$

From now on we consider $N/2$ sets of staggered fermions (with $N$ even) corresponding to $N$ flavors of Dirac fermions in the continuum limit. We suppress the additional flavor index. In the following expressions summation over all flavors is understood. In momentum space the constant auxiliary field is $\Phi(k) = (4\pi)^2 \Phi_0 \delta(k)$ and the Yukawa term takes the form

$$\begin{aligned} Y[\bar{\chi}, \chi, \Phi_0] &= \frac{1}{(4\pi)^2} \int_{\bar{B}} d^2k \, \bar{\chi}(-k) \sigma(k, -k) \chi(k) \Phi_0 \\ &= \frac{1}{(2\pi)^2} \int_B d^2k \, \bar{\chi}(-k) \lambda(k) \mathbb{1} \chi(k). \end{aligned} \tag{4.2}$$

Here we have used that $\bar{\chi}(-k)$ and $\chi(k)$ are $2\pi$-periodic and we have defined

$$\lambda(k)\mathbb{1} = \frac{1}{2^2} \sum_i \sigma(k + 2\pi i, -k - 2\pi i) \Phi_0, \tag{4.3}$$

where the sum extends over integer vectors with components $i_\mu \in \{0, 1\}$. For the standard action one finds $\lambda(k) = \Phi_0$. It takes the form

$$\begin{aligned} S[\bar{\chi}, \chi, \Phi_0] &= \frac{1}{(2\pi)^2} \int_B d^2k \, \bar{\chi}(-k)[\rho(k) + \lambda(k)\mathbb{1}]\chi(k) + \frac{1}{2}\Phi_0^2 V \\ &= \frac{1}{(2\pi)^2} \int_B d^2k \, \bar{\chi}(-k)[\rho^s(k) + \Phi_0 \mathbb{1}]\chi(k) + \frac{1}{2}\Phi_0^2 V, \end{aligned} \tag{4.4}$$

where $V$ is the space-time volume. The Yukawa term acts as a mass term of the fermion field. To find the fixed point action we proceed as in section 2. The calculation is exactly the same except that we now define $\alpha(k) + \beta(k)\mathbb{1} = [\rho(k) + \lambda(k)\mathbb{1}]^{-1}$. In particular, the recursion relation eq.(2.23) remains unchanged except that it now holds for $\tilde{\alpha}(k) + \beta(k)\mathbb{1}$. (We do not introduce $\tilde{\beta}(k)$ because it would be identical with $\beta(k)$.) Now the initial matrix elements characterizing the standard action are given by $\tilde{\alpha}_\mu(k) \sim ik_\mu/(k^2 + \Phi_0^2)$ and $\beta(k) \sim \Phi_0/(k^2 + \Phi_0^2)$ for small momenta. Thus, at the fixed point one finds

$$\tilde{\alpha}^*_\mu(k) = i \sum_{l \in \mathbb{Z}^2} \frac{k_\mu + 2\pi l_\mu}{(k + 2\pi l)^2 + \Phi_0^2} (-1)^{l_\mu} \prod_\nu \left( \frac{2 \sin(k_\nu/2)}{k_\nu + 2\pi l_\nu} \right)^2 + \frac{9}{8c} i \hat{k}_\mu,$$

$$\beta^*(k) = \sum_{l \in \mathbb{Z}^2} \frac{\Phi_0}{(k + 2\pi l)^2 + \Phi_0^2} \prod_\nu \left( \frac{2 \sin(k_\nu/2)}{k_\nu + 2\pi l_\nu} \right)^2. \quad (4.5)$$

Next we compare this result with the weak field approximation of the previous section. To $O(\Phi_0)$, $\alpha^*_\mu(k)$ is identical with the one of the free theory in eq.(2.26). We define $\tilde{\lambda}^*(k)\mathbb{1} = \tilde{\alpha}^*(k)\lambda^*(k)\mathbb{1}\tilde{\alpha}^*(k)$ and we find

$$\tilde{\lambda}^*(k) = -\beta^*(k) + O(\Phi_0^2) = - \sum_{l \in \mathbb{Z}^2} \frac{\Phi_0}{(k + 2\pi l)^2} \prod_\nu \left( \frac{2 \sin(k_\nu/2)}{k_\nu + 2\pi l_\nu} \right)^2 + O(\Phi_0^2). \quad (4.6)$$

On the other hand, analogous to eq.(4.3) one expects

$$\tilde{\lambda}^*(k)\mathbb{1} = \frac{1}{2^2} \sum_i \tilde{\sigma}^*(k + 2\pi i, -k - 2\pi i) \Phi_0. \quad (4.7)$$

From eq.(3.19) one obtains $\tilde{\sigma}^*_3(k, -k) = 0$ and

$$\begin{aligned}
\tilde{\sigma}^*_0(k, -k) &= \sum_{l,m \in \mathbb{Z}^2} \sum_i \frac{(k_\mu + 4\pi l_\mu + 2\pi i_\mu)(-k_\mu + 4\pi m_\mu + 2\pi i_\mu)}{(k + 4\pi l + 2\pi i)^2(-k + 4\pi m + 2\pi i)^2} \\
&\quad \times \prod_\nu (-1)^{l_\nu + m_\nu + i_\nu} \frac{2\sin(k_\nu/2)}{k_\nu + 4\pi l_\nu + 2\pi i_\nu} \frac{2\sin(-k_\nu/2)}{-k_\nu + 4\pi m_\nu + 2\pi i_\nu} \delta_{m_\nu, -l_\nu - i_\nu} \\
&= - \sum_{l \in \mathbb{Z}^2} \sum_i \frac{1}{(k + 4\pi l + 2\pi i)^2} \prod_\nu \left( \frac{2\sin(k_\nu/2)}{k_\nu + 4\pi l_\nu + 2\pi i_\nu} \right)^2 \\
&= - \sum_{l \in \mathbb{Z}^2} \frac{1}{(k + 2\pi l)^2} \prod_\nu \left( \frac{2\sin(k_\nu/2)}{k_\nu + 2\pi l_\nu} \right)^2 \quad (4.8)
\end{aligned}$$

such that indeed

$$\frac{1}{2^2} \sum_i \tilde{\sigma}^*(k + 2\pi i, -k - 2\pi i) = - \sum_{l \in \mathbb{Z}^2} \frac{1}{(k + 2\pi l)^2} \prod_\nu \left( \frac{2\sin(k_\nu/2)}{k_\nu + 2\pi l_\nu} \right)^2 \mathbb{1}. \quad (4.9)$$

In the next step we include the fluctuations of the nonzero modes. In the large $N$ limit they are small and can be treated by the small field approximation. Later we will be interested only in the fermion mass and in the chiral condensate. To leading order in $1/N$ these quantities don't pick up contributions from the nonzero modes. However, when one wants to compute e.g. the mass of the fermion-antifermion bound state (which we don't do in this paper), the dynamics of the nonzero modes enters at the lowest nontrivial order in $1/N$. We parametrize the Yukawa coupling as in eq.(3.10). However, now $\Phi(k)$ does not contain the zero mode $\Phi_0$ that was taken into account separately. The calculation of the fixed point action is very similar to the one in section 3. The only difference is that now $\tilde{\sigma}(p,q) = D(-p)^{1/2}[\alpha(p) + \beta(p)\mathbb{1}]\sigma(p,q)[\alpha(-q) + \beta(-q)\mathbb{1}]D(-q)^{1/2}$ which has the structure

$$\tilde{\sigma}(p,q) = \begin{pmatrix} \tilde{\sigma}_0(p,q) & \tilde{\sigma}_1(p,q) & \tilde{\sigma}_2(p,q) & -\tilde{\sigma}_3(p,q) \\ \tilde{\sigma}_1(p,q) & \tilde{\sigma}_0(p,q) & \tilde{\sigma}_3(p,q) & -\tilde{\sigma}_2(p,q) \\ \tilde{\sigma}_2(p,q) & -\tilde{\sigma}_3(p,q) & \tilde{\sigma}_0(p,q) & \tilde{\sigma}_1(p,q) \\ \tilde{\sigma}_3(p,q) & -\tilde{\sigma}_2(p,q) & \tilde{\sigma}_1(p,q) & \tilde{\sigma}_0(p,q) \end{pmatrix}. \quad (4.10)$$

The recursion is now started with

$$\tilde{\sigma}_0(p,q) = \frac{\hat{p}_\mu \hat{q}_\mu + \Phi_0^2}{(\hat{p}^2 + \Phi_0^2)(\hat{q}^2 + \Phi_0^2)} \prod_\nu \cos(\frac{p_\nu + q_\nu}{4}),$$

$$\tilde{\sigma}_\mu(p,q) = i\frac{(\hat{p}_\mu - \hat{q}_\mu)\Phi_0}{(\hat{p}^2 + \Phi_0^2)(\hat{q}^2 + \Phi_0^2)} \prod_\nu \cos(\frac{p_\nu + q_\nu}{4}),$$

$$\tilde{\sigma}_3(p,q) = \frac{\epsilon_{\mu\rho}\hat{p}_\mu \hat{q}_\rho}{(\hat{p}^2 + \Phi_0^2)(\hat{q}^2 + \Phi_0^2)} \prod_\nu \cos(\frac{p_\nu + q_\nu}{4}). \quad (4.11)$$

At the fixed point this leads to

$$\tilde{\sigma}_0^*(p,q) = \sum_{l,m \in \mathbb{Z}^2} \sum_i \frac{(p_\mu + 4\pi l_\mu + 2\pi i_\mu)(q_\mu + 4\pi m_\mu + 2\pi i_\mu) + \Phi_0^2}{[(p + 4\pi l + 2\pi i)^2 + \Phi_0^2][(q + 4\pi m + 2\pi i)^2 + \Phi_0^2]}$$

$$\times \prod_\nu (-1)^{l_\nu + m_\nu + i_\nu} \frac{2\sin(p_\nu/2)}{p_\nu + 4\pi l_\nu + 2\pi i_\nu} \frac{2\sin(q_\nu/2)}{q_\nu + 4\pi m_\nu + 2\pi i_\nu}$$

$$\times \frac{2\sin((p_\nu + q_\nu)/2)}{p_\nu + 4\pi l_\nu + q_\nu + 4\pi m_\nu + 4\pi i_\nu},$$

$$\tilde{\sigma}_\mu^*(p,q) = i\sum_{l,m \in \mathbb{Z}^2} \sum_i \frac{(p_\mu + 4\pi l_\mu - q_\mu - 4\pi m_\mu)\Phi_0(-1)^{i_\mu}}{[(p + 4\pi l + 2\pi i)^2 + \Phi_0^2][(q + 4\pi m + 2\pi i)^2 + \Phi_0^2]}$$

$$\times \prod_\nu (-1)^{l_\nu + m_\nu + i_\nu} \frac{2\sin(p_\nu/2)}{p_\nu + 4\pi l_\nu + 2\pi i_\nu} \frac{2\sin(q_\nu/2)}{q_\nu + 4\pi m_\nu + 2\pi i_\nu}$$

$$\times \frac{2\sin((p_\nu + q_\nu)/2)}{p_\nu + 4\pi l_\nu + q_\nu + 4\pi m_\nu + 4\pi i_\nu},$$

$$\tilde{\sigma}_3^*(p,q) = \sum_{l,m \in \mathbb{Z}^2} \sum_i \frac{\epsilon_{\mu\rho}(p_\mu + 4\pi l_\mu + 2\pi i_\mu)(q_\rho + 4\pi m_\rho + 2\pi i_\rho)}{[(p + 4\pi l + 2\pi i)^2 + \Phi_0^2][(q + 4\pi m + 2\pi i)^2 + \Phi_0^2]}$$

$$\times \prod_\nu (-1)^{l_\nu + m_\nu} \frac{2\sin(p_\nu/2)}{p_\nu + 4\pi l_\nu + 2\pi i_\nu} \frac{2\sin(q_\nu/2)}{q_\nu + 4\pi m_\nu + 2\pi i_\nu}$$

$$\times \frac{2\sin((p_\nu + q_\nu)/2)}{p_\nu + 4\pi l_\nu + q_\nu + 4\pi m_\nu + 4\pi i_\nu}. \quad (4.12)$$

Finally, the fixed point action is given by

$$\begin{aligned}\frac{1}{G}S^*[\bar{\chi},\chi,\Phi] &= \frac{1}{G}\{\frac{1}{(2\pi)^2}\int_B d^2k\,\bar{\chi}(-k)[\rho^*(k)+\lambda^*(k)\mathbf{1}]\chi(k) \\ &+ \frac{1}{2}\Phi_0^2 V + \frac{1}{(4\pi)^2}\int_{\bar{B}}d^2k\,\frac{1}{2}\Phi(-k)\Phi(k) \\ &+ \frac{1}{(4\pi)^4}\int_{\bar{B}^2}d^2p\,d^2q\,\bar{\chi}(-p)\sigma^*(p,q)\chi(-q)\Phi(p+q)\}. \quad (4.13)\end{aligned}$$

Only the first two terms contribute at $N = \infty$. The other terms represent the leading order of the small field approximation for the nonzero modes. They are important only at higher orders in $1/N$. It should be noted that the small field expansion can be extended systematically to higher orders. The above expression for the fixed point action has been derived in the limit $G \to 0$. We like to point out that it is the analog of the perfect classical action of Hasenfratz and Niedermayer that they determined numerically in the $O(3)$ model. Finding the fixed point action is a saddle point problem. In fact, we have found the fixed point just by solving classical equations of motion. Although we used a path integral formulation, so far we have only looked for saddle points. In this sense the fixed point action is a classical action. Of course, the path integrals one encounters in the large $N$ limit are all saddle point problems, and solving the classical problem is equivalent to doing the full integral. This is the reason why the fixed point action is even a perfect quantum action, i.e. there are no cut-off effects even if we leave the critical surface. This is the case only in the large $N$ limit. For finite $N$ one would expect some cut-off effects at finite correlation lengths even with the fixed point action. Still, the fact that the perfect classical action is a perfect quantum action in the large $N$ limit may explain why it is practically perfect in the $O(3)$ model studied by Hasenfratz and Niedermayer.

Next we construct a perfect operator for the chiral condensate. The standard operator is given by

$$\sum_x \bar{\chi}_x \chi_x = \frac{1}{(2\pi)^2}\int_B d^2k\,\bar{\chi}(-k)\chi(k). \quad (4.14)$$

The chiral condensate is a relevant operator — it gets amplified under renormalization group transformations and hence it is driven away from the fixed point. To determine the perfect operator we introduce a small perturbation around the fixed point

$$S_j[\bar{\chi},\chi,\Phi] = S^*[\bar{\chi},\chi,\Phi] + jX[\bar{\chi},\chi,\Phi]. \quad (4.15)$$

The perfect operator $X^*[\bar{\chi},\chi,\Phi]$ is an eigenfunctional that reproduces itself under renormalization, i.e.

$$S'_j[\bar{\chi}',\chi',\Phi'] = S^{*'}[\bar{\chi}',\chi',\Phi']+jX^{*'}[\bar{\chi}',\chi',\Phi'] = S^*[\bar{\chi}',\chi',\Phi']+j\gamma X^*[\bar{\chi}',\chi',\Phi']+O(j^2). \quad (4.16)$$

For relevant operators $\gamma > 1$. In two dimensions the chiral condensate has dimension $d_{\bar{\chi}\chi} = 1$. Hence, the corresponding eigenvalue is $\gamma = 3^{d_{\bar{\chi}\chi}} = 3$. Later we will be interested in the vacuum value of the perfect operator. To leading order in $1/N$ this quantity gets contributions only from the zero mode of the scalar auxiliary field. Hence we may assume that the scalar field is a constant $\Phi_0$. Still, if one wants to compute e.g. the perfect correlation function of $\bar{\chi}\chi$ also the nonzero modes contribute at the lowest nontrivial order in $1/N$. It is straightforward (but somewhat tedious) to derive the nonzero mode contribution in the small field approximation. Here we restrict ourselves to the zero mode contribution to the chiral condensate which we parametrize as

$$X[\bar{\chi}, \chi, \Phi_0] = \frac{1}{(2\pi)^2} \int_B d^2k\, \bar{\chi}(-k)[\mu(k) + \nu(k)\mathbb{1}]\chi(k). \tag{4.17}$$

Here $\mu(k)$ is a matrix of the same structure as $\rho(k)$. In general both $\mu(k)$ and $\nu(k)$ are functions of $\Phi_0$. The standard operator is characterized by $\mu(k) = 0$ and $\nu(k) = 1$. To determine the perfect operator we proceed analogous to the perfect action. First we define $[\alpha(k)_j + \beta(k)_j\mathbb{1}]^{-1} = \rho(k) + \lambda(k)\mathbb{1} + j[\mu(k) + \nu(k)\mathbb{1}]$ as well as $\tilde{\alpha}(k)_j = D(-k)^{1/2}\alpha(k)_j D(k)^{1/2}$ and $\tilde{\mu}(k) = D(-k)^{1/2}\mu(k)D(k)^{1/2}$. Then the recursion relation eq.(2.23) holds for $\tilde{\alpha}(k)_j + \beta(k)_j\mathbb{1}$, except that now

$$[\tilde{\alpha}'(k)_j + \beta'(k)_j\mathbb{1}]^{-1} = \tilde{\rho}'(k) + \lambda'(k)\mathbb{1} + j\gamma[\tilde{\mu}'(k) + \nu'(k)\mathbb{1}] + O(j^2). \tag{4.18}$$

Iterating eq.(2.23) one finds after infinitely many steps

$$\begin{aligned}
\tilde{\alpha}^*_\mu(k)_j &= \tilde{\alpha}^*_\mu(k) + j \lim_{n\to\infty} 3^n \partial_{\Phi_0}\tilde{\alpha}^*_\mu(k) + O(j^2), \\
\beta^*(k)_j &= \beta^*(k) + j \lim_{n\to\infty} 3^n \partial_{\Phi_0}\beta^*(k) + O(j^2).
\end{aligned} \tag{4.19}$$

This confirms $\gamma = 3$ and it allows to identify the perfect operator that is characterized by

$$\begin{aligned}
\tilde{\mu}^*_\mu(k) &= \partial_{\Phi_0}\left(\frac{\tilde{\alpha}^*_\mu(k)}{\tilde{\alpha}^*_\nu(-k)\tilde{\alpha}^*_\nu(k) + \beta^*(-k)\beta^*(k)}\right), \\
\nu^*(k) &= \partial_{\Phi_0}\left(\frac{\beta^*(k)}{\tilde{\alpha}^*_\nu(-k)\tilde{\alpha}^*_\nu(k) + \beta^*(-k)\beta^*(k)}\right).
\end{aligned} \tag{4.20}$$

Note that $\tilde{\alpha}^*_\mu(k)$ is odd in $k_\mu$ and even in the other components $k_\nu, \nu \neq \mu$, while $\beta^*(k)$ is even in all components of the momentum.

How local is the perfect operator for the chiral condensate? Here we answer this question only in the small field approximation omitting $O(\Phi_0)$ contributions, and we consider again effectively 1-dimensional configurations. From eq.(4.5) we obtain $\beta^*(k) = O(\Phi_0)$, $\partial_{\Phi_0}\tilde{\alpha}^*_\mu(k) = O(\Phi_0)$ and

$$\partial_{\Phi_0}\beta^*(k_1, 0) = \sum_{l_1 \in \mathbb{Z}} \frac{1}{(k_1 + 2\pi l_1)^2}\left(\frac{2\sin(k_1/2)}{k_1 + 2\pi l_1}\right)^2 + O(\Phi_0^2) = \frac{1}{\hat{k}_1^2} - \frac{1}{6} + O(\Phi_0^2). \tag{4.21}$$

Using eq.(4.20) together with eq.(2.27) for $c = 9$ one finds $\mu_\mu^*(k) = O(\Phi_0)$ and

$$\nu^*(k_1, 0) = 1 - \frac{1}{6}\hat{k}_1^2 + O(\Phi_0^2). \tag{4.22}$$

In coordinate space this implies for the perfect chiral condensate in the small field approximation

$$(\bar{\chi}\chi)_x = \frac{2}{3}\bar{\chi}_x\chi_x + \frac{1}{12}(\bar{\chi}_x\chi_{x+\hat{1}} + \bar{\chi}_{x+\hat{1}}\chi_x + \bar{\chi}_x\chi_{x-\hat{1}} + \bar{\chi}_{x-\hat{1}}\chi_x), \tag{4.23}$$

which is extremely local. For general 2-dimensional configurations the perfect chiral condensate not only contains nearest neighbor contributions, but still it is very local.

## 5 Solving the Gross-Neveu model at large $N$

In this section we solve the Gross-Neveu model at large $N$ using the full and small field fixed point actions as well as the standard action. For an explanation of the large $N$ technique we refer to Coleman's book [16]. We compute the energy spectrum of fermionic 1-particle states and the chiral condensate. This shows that the full fixed point action is a perfect quantum action. Up to now we were always on the critical surface, i.e. at infinite correlation length. Now we will use the perfect classical (fixed point) action away from the critical surface in a region where quantum fluctuations are present.

To solve the model we first compute the effective potential $V_{eff}(\Phi_0)$ for constant auxiliary scalar fields $\Phi_0$

$$\exp(-V_{eff}(\Phi_0)V) = \int \mathcal{D}\bar{\chi}\mathcal{D}\chi \exp(-\frac{1}{G}S[\bar{\chi}, \chi, \Phi_0]) = \det\mathcal{M}(\Phi_0)^{N/2}\exp(-\frac{1}{2G}\Phi_0^2 V), \tag{5.1}$$

where $V$ is the space-time volume, $\mathcal{M}(\Phi_0)$ is the fermion matrix of a single set of staggered fermions in the constant background field $\Phi_0$, and the power $N/2$ accounts for the various flavors. Up to now $S[\bar{\chi}, \chi, \Phi_0]$ is still general. We parametrize it as before

$$S[\bar{\chi}, \chi, \Phi_0] = \frac{1}{(2\pi)^2}\int_B d^2k\,\bar{\chi}(-k)[\rho(k) + \lambda(k)\mathbb{1}]\chi(k) + \frac{1}{2}\Phi_0^2 V, \tag{5.2}$$

where $\rho(k)$ and $\lambda(k)$ are functions of $\Phi_0$. One obtains

$$V_{eff}(\Phi_0) = \frac{1}{2G}\Phi_0^2 - \frac{N}{2}\frac{1}{(2\pi)^2}\int_B d^2k \log\det\frac{1}{G}[\rho(k) + \lambda(k)\mathbb{1}]. \tag{5.3}$$

The minimum of the effective potential is given by

$$\partial_{\Phi_0}V_{eff}(\Phi_0) = \frac{\Phi_0}{G} - \frac{N}{2}\frac{1}{(2\pi)^2}\int_B d^2k\frac{\partial_{\Phi_0}\det[\tilde{\rho}(k) + \lambda(k)\mathbb{1}]}{\det[\tilde{\rho}(k) + \lambda(k)\mathbb{1}]} = 0. \tag{5.4}$$

Here we have used $\det[\rho(k) + \lambda(k)\mathbb{1}] = \det[\tilde{\rho}(k) + \lambda(k)\mathbb{1}]$. Eq.(5.4) is an implicit equation for $\Phi_0$, known as the gap equation. It determines the relation between the bare coupling $G$ and the vacuum value of the auxiliary field $\Phi_0$. For the standard action $\tilde{\rho}_\mu(k) = -i\hat{k}_\mu$ and $\lambda(k) = \Phi_0$. Then the gap equation reduces to

$$\frac{1}{g} = \frac{1}{(2\pi)^2} \int_B d^2k \, \frac{2}{\hat{k}^2 + \Phi_0^2}. \tag{5.5}$$

We recall that the large $N$ limit is taken such that the coupling $g = GN$ is kept fixed. The gap equation for the fixed point action is given by

$$\frac{\Phi_0}{g} = -\frac{1}{(2\pi)^2} \int_B d^2k \, \frac{2[\tilde{\alpha}_\nu^*(-k)\partial_{\Phi_0}\tilde{\alpha}_\nu^*(k) + \beta^*(-k)\partial_{\Phi_0}\beta^*(k)]}{\tilde{\alpha}_\nu^*(-k)\tilde{\alpha}_\nu^*(k) + \beta^*(-k)\beta^*(k)}. \tag{5.6}$$

Next we investigate the spectrum of the fermionic 1-particle states. For purely technical reasons related to the proper normalization of momentum eigenstates the system is put in a finite spatial volume of size $L$ with periodic boundary conditions. We consider the correlation function of a fermionic operator with definite spatial momentum $k_1 = 2\pi n_1/L \in ]-\pi, \pi]$, $n_1 \in \mathbf{Z}$

$$\chi(k_1)_{x_2} = \frac{1}{2\pi} \int_{-\pi}^{\pi} dk_2 \, \chi(k_1, k_2) \exp(ik_2 x_2) \tag{5.7}$$

that creates a fermion at euclidean time $x_2$. The correlation function is given by

$$\langle \bar{\chi}(-k_1)_0 \chi(k_1)_{x_2} \rangle = \frac{L}{2\pi} \int_{-\pi}^{\pi} dk_2 \, \frac{N}{2} \text{Tr} \mathcal{M}(\Phi_0)^{-1} \exp(ik_2 x_2). \tag{5.8}$$

Here the trace is over the pseudoflavor index of a single set of staggered fermions only. The inverse fermion matrix is given by $\mathcal{M}(\Phi_0)^{-1} = -G[\alpha(k) + \beta(k)\mathbb{1}]$. For the standard action we have $\beta(k) = \Phi_0/(\hat{k}^2 + \Phi_0^2)$ such that

$$\langle \bar{\chi}(-k_1)_0 \chi(k_1)_{x_2} \rangle = -\frac{L}{2\pi} \int_{-\pi}^{\pi} dk_2 \, \frac{2g\Phi_0 \exp(ik_2 x_2)}{\hat{k}_1^2 + \hat{k}_2^2 + \Phi_0^2} = C(k_1) \exp(-E(k_1)x_2). \tag{5.9}$$

As expected the correlation function decays exponentially, and the energy of the 1-fermion state is given by the pole of the integrand as $E(k_1) = -ik_2$ with

$$4\sinh^2(E(k_1)/2) = -4\sin^2(k_2/2) = -\hat{k}_2^2 = \hat{k}_1^2 + \Phi_0^2. \tag{5.10}$$

The fermion mass $m_f$ is the energy of the $k_1 = 0$ state, i.e.

$$2\sinh(m_f/2) = \Phi_0. \tag{5.11}$$

For small $m_f$ and $k_1$ the momentum dispersion relation turns into the one of the continuum theory. However, there are $O(a^2)$ corrections to it. It is interesting to note that all cut-off effects are $O(a^2)$, while one would naively expect them to be of $O(a)$ in a fermionic model. Since we have not introduced an explicit chiral symmetry breaking fermion mass term, the action is $\mathbf{Z}(2)$ chirally invariant. This symmetry prevents the generation of irrelevant operators of dimension 3, which would induce $O(a)$ scaling violations. For the same reason the standard action is in this case equivalent to an $O(a)$ Symanzik improved action. Hence, it is of the same quality as the Sheikholeslami-Wohlert action for QCD. For the small field fixed point action one finds

$$2\sinh(m_f/2) = \frac{1 - \sqrt{1 - 2\Phi_0^2/3}}{\Phi_0/3}, \tag{5.12}$$

and again there are $O(a^2)$ corrections to the continuum dispersion relation. The small field fixed point action is definitely not a perfect action. Finally we consider the full fixed point action. Then using eq.(4.5) the eq.(5.8) turns into

$$\begin{aligned}
\langle \bar{\chi}(-k_1)_0 \chi(k_1)_{x_2} \rangle &= -\frac{L}{2\pi} \int_{-\pi}^{\pi} dk_2 \sum_{l \in \mathbf{Z}^2} \frac{2g\Phi_0 \exp(ik_2 x_2)}{(k+2\pi l)^2 + \Phi_0^2} \prod_\nu \left(\frac{2\sin(k_\nu/2)}{k_\nu + 2\pi l_\nu}\right)^2 \\
&= -\frac{L}{2\pi} \int_{-\infty}^{\infty} dk_2 \sum_{l_1 \in \mathbf{Z}} \frac{2g\Phi_0 \exp(ik_2 x_2)}{(k_1+2\pi l_1)^2 + k_2^2 + \Phi_0^2} \left(\frac{2\sin(k_1/2)}{k_1 + 2\pi l_1}\right)^2 \left(\frac{2\sin(k_2/2)}{k_2}\right)^2 \\
&= \sum_{l_1 \in \mathbf{Z}} C(k_1 + 2\pi l_1) \exp(-E(k_1 + 2\pi l_1)x_2). \tag{5.13}
\end{aligned}$$

The sum over $l_2$ and the integration of $k_2$ over the Brillouin zone have combined to an integration over the momentum space of the continuum theory. The remaining summation over $l_1$ leads to infinitely many poles of the integrand, and hence to infinitely many states that contribute to the correlation function. Their energies are given by $E(k_1 + 2\pi l_1) = -ik_2$ with

$$E^2(k_1 + 2\pi l_1) = -k_2^2 = (k_1 + 2\pi l_1)^2 + \Phi_0^2. \tag{5.14}$$

This is the *exact* continuum dispersion relation for a particle of mass $m_f = \Phi_0$ with momentum $k_1 + 2\pi l_1$. Remarkably, this momentum is unlimited although the lattice momentum $k_1$ is restricted to the Brillouin zone $]-\pi, \pi]$. This means that the full fixed point action has exactly the same fermionic 1-particle spectrum as the continuum theory. There are no cut-off effects — the action is perfect. It is interesting to see how the perfect action can account for all momentum states of the continuum theory. Since it is a lattice action it is invariant only against translations by multiples of the lattice spacing. Hence, like for any other lattice action, using the discrete translation symmetry one can only select states which have a definite lattice momentum $k_1 \in ]-\pi, \pi]$. Still, for the fixed point action the other momentum states $k_1 + 2\pi l_1$ of the continuum theory show up, and they all contribute to the correlation function with the lattice momentum $k_1$ that was selected by symmetry. This is possible only because the fixed point action is extended in euclidean time. In principle, it extends over all time slices, but it falls off extremely fast, and it is hence practically local. Still, if one wants to catch even the highest states of the continuum theory, one must take the coupling of all time slices into account. In practice, e.g.

in a Monte-Carlo simulation, one would not like to keep the exponentially small tails of the fixed point action. When one cuts them off one does some harm to the continuum spectrum. However, this only affects the highest states, while the energies of low lying states are still practically perfect. The same remarks apply to rotation invariance. As a lattice action the perfect action is not spherically symmetric. Its symmetry is restricted to the discrete cubic rotation group. Still, the spectrum does not notice this — it is the exact one of the continuum theory. The lattice symmetry only allows us to construct operators that project on a given representation of the cubic rotation group. Still, all states of the continuum theory have some projection on these operators. Remarkably, when the fixed point action is used, their energy is perfect. Again, this is possible only because the action is extended in space. Cutting off its exponentially suppressed tail will do some harm, but it will mostly affect the high angular momentum states that one does not focus on in numerical studies.

Now let us consider the chiral condensate. Using the standard action together with the standard operator for the chiral condensate one finds

$$\langle \bar{\chi}\chi \rangle = \frac{1}{(2\pi)^2} \int_B d^2k \frac{N}{2} \text{Tr}\mathcal{M}(\Phi_0)^{-1} = -\frac{1}{(2\pi)^2} \int_B d^2k \frac{2g\Phi_0}{\hat{k}^2 + \Phi_0^2} = -\Phi_0. \qquad (5.15)$$

Here we have used the gap equation (5.5). The ratio $\langle \bar{\chi}\chi \rangle/m_f$ should scale, i.e. it should be cut-off independent in the continuum limit. Indeed

$$\langle \bar{\chi}\chi \rangle/m_f = -2\sinh(m_f/2)/m_f, \qquad (5.16)$$

which approaches the constant $-1$ in the continuum limit $\xi = 1/m_f \to \infty$. However, there are $O(a^2)$ cut-off effects as long as the correlation length $\xi$ is finite. Using the perfect action with the perfect operator for the chiral condensate one obtains

$$\begin{aligned}\langle \bar{\chi}\chi \rangle &= -\frac{1}{(2\pi)^2} \int_B d^2k \frac{g}{2} \text{Tr}\{[\mu^*(k) + \nu^*(k)\mathbb{1}][\alpha^*(k) + \beta^*(k)\mathbb{1}]\} \\ &= \frac{1}{(2\pi)^2} \int_B d^2k \frac{2g[\tilde{\alpha}_\nu^*(-k)\partial_{\Phi_0}\tilde{\alpha}_\nu^*(k) + \beta^*(-k)\partial_{\Phi_0}\beta^*(k)]}{\tilde{\alpha}_\nu^*(-k)\tilde{\alpha}_\nu^*(k) + \beta^*(-k)\beta^*(k)} = -\Phi_0. \end{aligned} \qquad (5.17)$$

Here we have used eqs.(4.20,5.6). Now $\langle \bar{\chi}\chi \rangle/m_f = -1$ even for arbitrarily small $\xi$. Thus, cut-off effects are completely eliminated. It should be noted that this does not happen automatically just by using the perfect action. Also the operator for the chiral condensate must be perfect. With the standard operator $\bar{\chi}_x\chi_x$ there would be cut-off effects even when the perfect action is used.

Finally, we investigate the *asymptotic* scaling behavior. For the standard action this has already been done in ref.[11]. Fig.4 depicts $m_f(g)$ for the perfect action and for the standard action. For the perfect action asymptotic scaling sets in earlier than for the standard action. In fact, for the perfect action the deviations from asymptotic scaling are only about 5 % for correlation lengths as small as 1. Note that this corresponds to 2 lattice units on the fine lattice (with spacing 1/2). Still, asymptotic scaling is not perfect for the perfect action. Fig.5 compares the $\beta$-function $\beta(g) = -m_f \partial_{m_f} g$ of the perfect and the standard action with the perturbative one obtained from eq.(1.5)

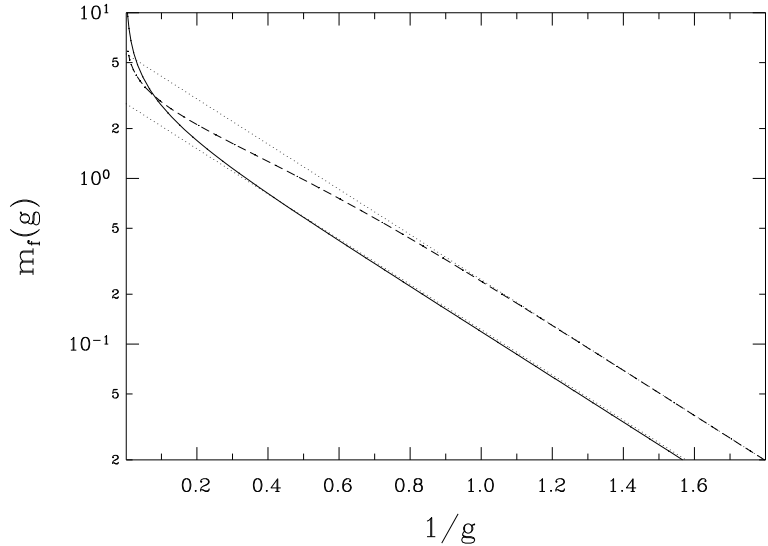

Figure 4: *Scaling behavior $m_f(g)$ of the perfect action (solid line) and of the standard action (dashed line). Asymptotic scaling with the perturbative $\beta$-function corresponds to the straight dotted lines.*

$$\beta(g) = -g^2/\pi. \tag{5.18}$$

Again, one sees that asymptotic scaling sets in earlier with the perfect action, but it is not perfect. Asymptotic scaling is not what one should ask for. What really matters is that the physics (dimensionless ratios of quantities in lattice units) are cut-off independent. This is the case for the perfect action. The fact that it does not show perfect asymptotic scaling is unphysical because it involves the bare coupling defined at the cut-off scale. Scaling, on the other hand, is a physical issue, and scaling is perfect for the fixed point action.

It may still seem mysterious why the fixed point action — which is a perfect classical action — is even perfect on the quantum level, i.e. why there are no cut-off effects even at arbitrarily small correlation length. To shed some light on this question we investigate the renormalized trajectory of theories with perfect quantum actions in the large $N$ limit. We want to show that the classical perfect action is in fact located on the trajectory. Let us say we are aiming at the perfect quantum action for the theory at some finite correlation length $\xi = 1/m_f$. We construct this action by performing $n$ exact renormalization group steps starting at a point arbitrarily close to (but not on) the critical surface, i.e. we start at a correlation length $3^n \xi$. We are interested in the limit $n \to \infty$ with $m_f = 1/\xi$ fixed. Of course, now we are at $g \neq 0$ and for finite $N$ performing a renormalization group step is no longer a saddle point problem. Instead one must perform the full path integral. At large $N$, however, we may again assume that only the zero mode $\Phi_0$ of the auxiliary scalar field has a large value, while the nonzero modes can be treated in the small field approximation. Then again *all* path integrals are saddle point problems, and their result is identical with the classical value. Thus, an infinite number of renormalization group steps leading us down the renormalized trajectory results again in the perfect classical action, except that now $g \neq 0$. Clearly, this is the case only in the large $N$ limit. At finite $N$ one expects the perfect classical action to have some cut-off effects at small correlation lengths.

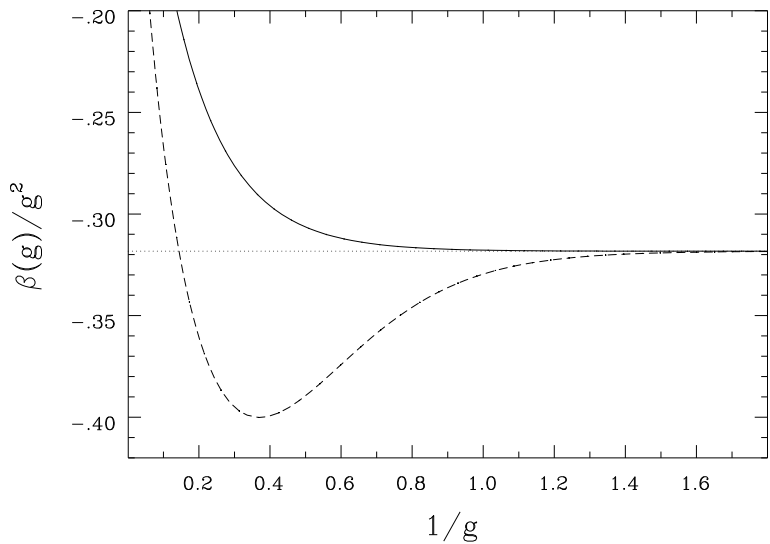

Figure 5: *The $\beta$-function (more precisely $\beta(g)/g^2$ versus $1/g$) for the perfect action (solid line) and for the standard action (dashed line). Asymptotic scaling with the perturbative $\beta$-function corresponds to the dotted line.*

# 6   Conclusions

The main purpose of the present paper was to investigate fermionic fixed point actions in a simple setting, and to understand better the success of Hasenfratz's and Niedermayer's perfect action approach in an analytic manner. This confined us to the large $N$ limit, because only there we could construct the fixed point action analytically. It turned out that the perfect classical (fixed point) action is in fact also a perfect quantum action. Cut-off effects are completely eliminated even at arbitrarily small correlation lengths. Certainly, this is the case only in the large $N$ limit. It is related to the fact that then all path integrals are saddle point problems and classical methods already give the full answer. Still, we believe that this may explain why the classical fixed point action worked so well in the $O(3)$ model. The fact that the perfect classical action is perfect on the quantum level at large $N$ may imply that it is still very good for $N = 3$. This argument assumes, of course, that our result is also relevant for the $O(N)$ model, but we are confident that this is actually the case. It would be nice if one could argue that the perfect classical action should then work equally well for nonabelian gauge theories. Unfortunately, already the large $N$ limit of nonabelian gauge theories is so complicated that a similar analytic study is beyond our technical abilities. Finally, we like to comment once more on scaling versus asymptotic scaling. As our results show (and as various people have realized before) one should ask for *scaling* not for asymptotic scaling. In particular, the perfect action represents exact continuum physics although the bare lattice coupling does not exactly follow the perturbative $\beta$-function. Still, for the perfect action asymptotic scaling sets in earlier than for the standard action.

Although our real interest lies in QCD we believe that it may be helpful to gain further experience with fixed point actions for fermions in simpler settings like the Gross-Neveu model. For example, it would be interesting to investigate the fixed point action for a single set of staggered fermions ($N = 2$). Definitely, this is a numerical problem. Still, many results of the present paper are relevant for such an investigation. First, the locality of the fixed point action for free fermions has been optimized analytically resulting in $c = 9$ for the free parameter in the renormalization group transformation. This parameter should also be used when one looks for the perfect action at $N = 2$. Second, the small field fixed point action was derived at finite $N$. This is certainly the first step towards the full fixed point action. However, there is a lesson that we learnt: the small field fixed point action is not a perfect action. Hence, one should not rely on that approximation. Finally, the knowledge of the large $N$ fixed point action may help to find a good parametrization for the perfect action at finite $N$. It is also important to explore fermionic fixed point actions in numerical simulations. In particular, it will be interesting to investigate how well standard numerical techniques like the Hybrid-Monte Carlo algorithm work in this case.

It would be interesting to extend the large $N$ analysis by systematically calculating $1/N$ corrections to the fixed point action. This could be done not only for the Gross-Neveu model but also for $O(N)$ and $CP(N)$ models. Of course, the real challenge is to develop perfect actions for the theory of strong interactions.

# Acknowledgements

We like to thank P. Hasenfratz, J. Jersák and F. Niedermayer for very interesting discussions, and J. Smit for a helpful remark. One of us (W.B.) likes to thank the Centro Brasileiro de Pesquisas Fisicas (Brazil) for warm hospitality.